\documentclass[pre,superscriptaddress,floatfix,longbibliography,twocolumn]{revtex4-1}

\usepackage{graphics}
\usepackage{graphicx}
\usepackage{amsmath}
\usepackage{amssymb,xcolor}
\usepackage{textcomp}
\usepackage{siunitx}
\usepackage{soul}
\newcommand{\CFL}[1]{\textcolor{teal}{#1}}

\newcommand{\textin}[1]{\mbox{\scriptsize{#1}}}

\definecolor{grisclair}{rgb}{0.6,0.6,0.6}

\newcommand{\beq}{\begin{equation}}
\newcommand{\ee}{\end{equation}}

\begin{document}

\title{Steady cone-jet mode of electrospray for single-cell deposition}
\author{D. Fern\'andez-Mart\'{\i}nez}
\email{danielfm@unex.es}
\address{Depto.\ de Ingenier\'{\i}a Mec\'anica, Energ\'etica y de los Materiales and\\ 
Instituto de Computaci\'on Cient\'{\i}fica Avanzada (ICCAEx),\\
Universidad de Extremadura, E-06006 Badajoz, Spain}
\author{C. Ferrera}
\address{Depto.\ de Ingenier\'{\i}a Mec\'anica, Energ\'etica y de los Materiales and\\ 
Instituto de Computaci\'on Cient\'{\i}fica Avanzada (ICCAEx),\\
Universidad de Extremadura, E-06006 Badajoz, Spain}
\author{J. M. Montanero}
\address{Depto.\ de Ingenier\'{\i}a Mec\'anica, Energ\'etica y de los Materiales and\\ 
Instituto de Computaci\'on Cient\'{\i}fica Avanzada (ICCAEx),\\
Universidad de Extremadura, E-06006 Badajoz, Spain}
\author{L. Mendoza-Cerezo}
\address{Depto.\ de Expresi\'on Gr\'afica. Universidad de Extremadura, E-06006 Badajoz, Spain}
\address{Depto.\ de Bioqu\'{\i}mica. Universidad de Extremadura, E-06006 Badajoz, Spain}
\author{J. M. Rodr\'{\i}guez-Rego}
\address{Depto.\ de Expresi\'on Gr\'afica. Universidad de Extremadura, E-06006 Badajoz, Spain}

\begin{abstract}
We propose using the electrospray cone-jet mode operated near its minimum-flow-rate stability limit for single-cell deposition. Because the jet is much thinner than the cells themselves, individual cells can be clearly visualized or detected during deposition. At such low flow rates, individual cells can be placed at distinct, user-defined locations, even at relatively high cell concentrations. In this sense, our approach provides a spatial resolution at the scale of a single cell. We demonstrate the method's capabilities by depositing cells onto a millimeter-scale droplet of a standard cell-culture medium. Cell viability assays indicate that many cells maintain membrane integrity after exposure to the electrosprayed liquid, suggesting that most damage is reversible. 
\end{abstract}

\maketitle

\section{Introduction}

Bioprinting enables the fabrication of living constructs, such as tissues and organs \citep{KKPA25,GCL24}. These constructs have the potential to address limitations in transplantation, traumatic injury repair, and congenital defect correction, which remain clinically unresolved due to limited donor availability and immune rejection \citep{WCLB25, HSA25}. Bioprinting can also assist surgeons in preoperative planning \citep{ZHBSEQMYK13},  enable cosmetic and drug testing \citep{DRGPSEO25}, and support a broad spectrum of biomedical research areas \citep{DPHB21,NYPS22,KKPA25}.

Current bioprinting strategies can replicate macroscopic structures but still lack the ability to position individual cells with micrometer accuracy \citep{C07,ZA20,ZMHMLLX22}. Such precision is essential to recreate the structural and biochemical gradients of native tissues and organs \citep{B21b,DPHB21,MBC24,KKPA25} and to control cell density and spatial arrangement when studying cell-cell \citep{ZWWZ22} and cell-scaffold \citep{GCL24} interactions, which remain difficult to elucidate using conventional methods \citep{JHKJK20,DPHB21}. 

Single-cell manipulation is also an accessible alternative to isolation approaches such as limiting dilution, micromanipulation, flow cytometry, or microfluidic trapping systems \citep{RGYSYR25,LW23,MZSJ25}, many of which are costly, labor-intensive, or may compromise cell integrity. Single-cell manipulation is crucial given the inherent heterogeneity within cell populations \citep{ZWWCW20,RGYSYR25,LW23}, and enables the study of cytobiological processes at the single-cell resolution \citep{LFWKZ21,KSMTIK21}. Moreover, it facilitates the incorporation of biomolecules and bioactive reagents \citep{WCLB25,HSA25} that enhance biocompatibility and cell differentiation, adhesion, and proliferation \citep{DHK13, KXLCW18}. Single-cell bioprinting can also incorporate microchannels or endothelial cells into engineered tissues \citep{GFLH20}, thereby enhancing nutrient delivery and addressing the lack of vascularization in conventional monolayer cultures and spheroids. Bioprinted biosensors at the cellular scale allow the assessment of drug-induced responses and dynamic changes in cellular behavior \citep{DRGPSEO25,DMU25}.

The required resolution for these applications can be achieved through contact \citep{QWHLCZ21}, laser-based \citep{AGENES21,BGKMMLL24}, inkjet, and photopolymerization bioprinting \citep{GCL24}. However, contact bioprinting is limited to two-dimensional cell arrays, and the stamping process can damage or contaminate the printed pattern \citep{ZWWZ22}. Laser-based bioprinting is expensive, can induce thermal damage to cells, and cannot print thick structures \citep{DRGPSEO25}. Moreover, compatible printers and biogels are scarce, and exhibit lower processing throughput than other methods due to their complexity \citep{AGENES21,GFLH20,GCL24}. Inkjet bioprinting can produce 3D tissues at a single-cell level \citep{KEZ21}, but it is limited to low-viscosity or low-cell-concentration bioinks owing to nozzle clogging \citep{GFLH20,GCL24,DRGPSEO25}.  Finally, stereolithography, digital light projection, and two-photon polymerization are expected to become dominant in the future because they offer very high resolution \citep{GFLH20}. Nevertheless, they remain under development, are very slow, and utilize limited, wasteful, and expensive bioresins \citep{KAPFGLXS24}.

Electrohydrodynamic printing techniques, such as electrospray, have the potential to overcome the drawbacks mentioned above. The steady cone-jet mode of electrospray enables the continuous ejection of a carrying jet much thinner than the feeding nozzle (capillary) \citep{G04a,M24}. Electric charges accumulate at the free surface of an electrified cone (the Taylor cone) attached to the nozzle. This produces interfacial Maxwell stresses that accelerate the liquid toward the cone tip. The liquid is evacuated through a thin, electrified jet. For most low-conductivity Newtonian liquids, the jet velocity essentially depends only on the liquid properties, implying that the diameter scales as the square root of the injected flow rate \citep{G04a,M24}. Therefore, the jet diameter can be reduced by decreasing the flow rate until the minimum-flow-rate stability limit \citep{G04a,M24,PRHGM18}. The minimum flow rate in electrospray is typically less than 100 $\mu$l/h, resulting in very low shear stress in the feeding capillary and thereby preventing cell damage. 

The liquid used in the steady cone-jet mode of bioprinting must meet two conditions: (i) its chemical composition must allow cell viability, and (ii) the viscosity and electrical conductivity must enable the stable cone-jet mode. These two conditions are somewhat antagonistic. Chemical products that improve cell survival increase electrical conductivity to values incompatible with Taylor cone stability \citep{GJF20,JAGANMJ24}. For this reason, there have been very few experimental realizations of the steady cone-jet mode for bioprinting. None of them enables single-cell deposition.  

\citet{JT06} used the coaxial electrospray configuration to eject a highly concentrated biological solution with a level of resolution of 50 $\mu$m. In this configuration, an outer liquid coflows with the inner liquid carrying the cells. The driving Maxwell stress is applied on the outermost surface. Therefore, the outer liquid must comply with electrospray requirements, while the inner liquid composition can be adjusted to ensure cell viability.  \citet{KAIMJ08} combined electrospray and flow focusing to handle a wide range of cells for deposition.

Under specific parameter conditions, viscoelasticity stabilizes the Taylor cone, and fibers thinner than the feeding capillary can be electrospun. However, these fibers commonly exhibit whipping instability, which prevents controlled deposition. \citet{CLBNRSDPIIM21} engineered two hydrogel bioinks, based on gelatin and silk fibroin, that enabled the controlled printing of fibers emitted by the steady cone-jet mode of electrospray. A summary of electrohydrodynamic techniques using polymer and hydrogel solutions can be found in  Ref.\ \citep{CLBNRSDPIIM21}.  

In this paper, we propose using the simple quasi-Newtonian cone-jet mode configuration, operating near the minimum-flow-rate stability limit, for cell-by-cell deposition into a cell-culture medium. The jet is much thinner than the cells, allowing individual cells to be visualized. The flow rate is so low that individual cells can be placed at distinct, user-defined locations, even at relatively high cell concentrations. First, we show the stability and robustness of this technique. Then, we demonstrate its performance by depositing cells onto a millimeter-sized droplet of a commonly used cell-culture medium. Our results indicate that each cell can be deposited at a different location, even for the moderately large cell concentration $c=2\times 10^6$ cells/ml used in our experiments. The technique can be used in bioprinting to achieve the maximum possible spatial resolution, i.e., the cell size.

\section{Methods}

\subsection{Liquids}

The electrosprayed liquid was a concentrated suspension of MCF-7 human breast adenocarcinoma cells in aqueous polyethylene glycol (PEG) 35 kDa at 10\% (w/v). The cell concentration was $c=2\times 10^6$ cells/ml. This PEG solution was prepared by adding the polymer to ultrapure type I water to minimize the electrical conductivity. This was done in a sterile container by manual agitation via continuous tube inversion until complete solubilization was achieved. The solution was subsequently sterilized by filtration using a 0.22 $\mu$m filter. The MCF-7 cell line was obtained from the Biomedical Research Support Services of the University of Extremadura and maintained in high-glucose Dulbecco's Modified Eagle Medium (DMEM) GlutaMAX (Gibco) supplemented with 10 fetal bovine serum (Gibco) and 1\% penicillin-streptomycin (Thermo Fisher Scientific). Cells were cultured at 37 $^{\circ}$C, 95\% humidity, and 5\% CO$_2$, with periodic medium changes until subconfluence was reached. Subculturing was performed by trypsinization using trypsin-EDTA (Gibco), and cells were maintained in 75 cm$^2$ culture flasks. 

The physical parameters of the PEG solution were: density $\rho=1015$ kg/m$^3$, viscosity $\mu=23.4$ mPa$\cdot$s, surface tension $\sigma=59.9$ mN/m, electrical conductivity $K=6.1$ mS/m, and electric permittivity $\varepsilon=77 \varepsilon_o$, where $\varepsilon_o$ is the vacuum permittivity. Shear rheological measurements showed no shear-thinning behavior across the entire range of shear rates investigated. Extensional rheometry was performed using the slow retraction method \citep{SVSMA17}. The results indicate that the extensional relaxation time $\lambda_s$ is below 30 $\mu$s, implying that the liquid is weakly viscoelastic (see the Supplemental Material). The osmolality of the PEG solution was 30 mOsm.

The electrosprayed liquid was deposited onto DMEM supplemented with 10\% (v/v) fetal bovine serum (FBS) and 5\% (v/v) penicillin–streptomycin (Pen/Strep), a commonly used cell culture medium. The physical parameters of this liquid were: density $\rho=1020$ kg/m$^3$, viscosity $\mu=1.1$ mPa$\cdot$s, surface tension $\sigma=52.2$ mN/m, electrical conductivity $K=1.5$ S/m, and electric permittivity $\varepsilon=77 \varepsilon_o$. This liquid cannot be electrosprayed due to its high electrical conductivity and viscosity. 

\subsection{Electrospray}
\label{sec21}

\begin{figure}
\begin{center}
\resizebox{0.45\textwidth}{!}{\includegraphics{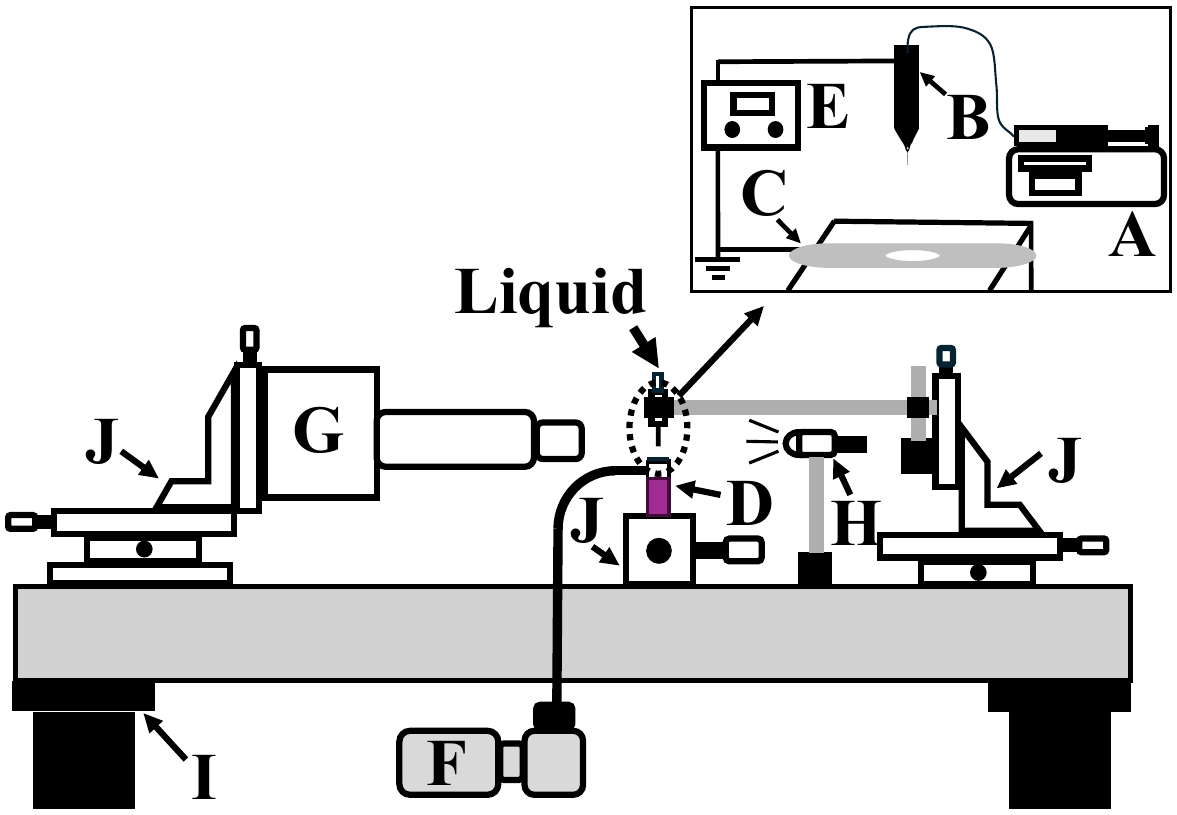}}
\end{center}
\vspace{0cm}
\caption{Sketch of the experimental setup to visualize the cone-jet mode of electrospray: syringe pump (A), feeding capillary (B), metallic plate (C), cubic chamber (D), DC high-voltage power supply (Bertan 205B-10R) (E), suction pump (F), high-speed camera (G), optical fiber (H), optical table (I), and (J) orientation system.}
\label{Exp_Sketch}
\end{figure}

We began our analysis by examining cell ejection in the steady cone–jet regime of electrospray. The experimental setup used in these experiments is shown schematically in Fig.\ \ref{Exp_Sketch}. The PEG solution was injected at a constant flow rate $Q$ by a syringe pump (Harvard Apparatus PHD 4400) (A) connected to a stepper motor through a cylindrical metallic capillary (B). The capillary was $D_i=150$ $\mu$m ($D_o=310$ $\mu$m) in inner (outer) diameter, and was located at an adjustable distance $H$ from a metallic plate (C). The plate had an orifice of diameter 800 $\mu$m in front of the capillary. The plate covered the upper face of a borosilicate cubic chamber (D). An electric potential $V$ was applied to the end of the feeding capillary via a DC high-voltage power supply (Bertan 205B-10R) (E). The metallic plate served as the ground electrode. A prescribed negative gauge pressure (approximately 10 mbar) was applied in the cubic chamber using a suction pump (F) to produce an air stream co-flowing with the jet across the plate orifice. The gas stream prevented liquid accumulation on the metallic plate but did not affect jet emission. The cubic chamber was partially filled with the cell-culture medium. The cells were placed in this liquid bath to assess their viability. We produced a stable cone-jet for $0.5\leq H\leq 1.3$ mm, $1.7\leq V\leq 3.6$ kV, and $0.03\leq Q\leq 0.15$ ml/h. The results of these experiments, presented in Sec.\ \ref{sec31}, were obtained for $H=0.5$ mm, $V=1.7$ kV, and $Q=0.1$ ml/h.

Digital images of the liquid jet were acquired at $10^4$ frames per second with an exposure time in the interval $3.9-12.5$ $\mu$s  using a high-speed camera (Photron FastCam Mini UX100) (G) equipped with optical lenses (Optem Fusion). The magnification was $0.5$ $\mu$m/pixel. The fluid configuration was illuminated from the back by cold white light delivered via an optical fiber (H). To check that the fluid configuration was axisymmetric, we also acquired images of the liquid meniscus using an auxiliary CCD camera (not shown in Fig.\ \ref{Exp_Sketch}) with an optical axis perpendicular to that of the primary camera. All these elements were mounted on an optical table provided with an antivibration isolation system (I).

\begin{figure}
\begin{center}
\resizebox{0.18\textwidth}{!}{\includegraphics{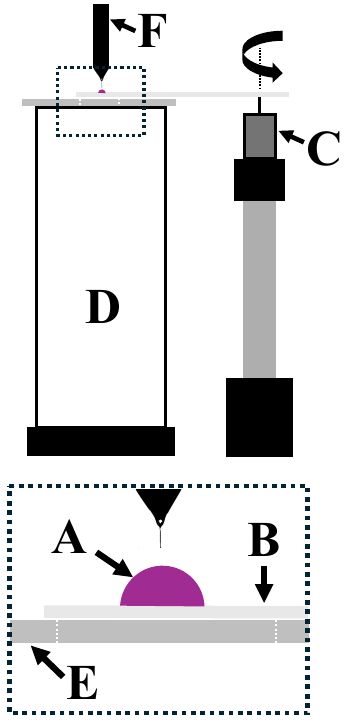}}
\end{center}
\vspace{0cm}
\caption{Sketch of the experimental setup to visualize cell-by-cell deposition: droplet of cell culture medium (A), glass slide (B), stepper motor (C), cubic chamber (D), metallic plate (E), and (F) feeding capillary.}
\label{Exp_Sketch2}
\end{figure}

In the second part of our study, we conducted experiments to demonstrate the capability of our method for single-cell deposition onto a droplet of the culture medium (Fig.\ \ref{Exp_Sketch2}). In this case, we proceeded as follows. First, we placed a millimeter-sized droplet of cell culture medium (A) on a glass slide (B). The glass was hydrophilic, with a roughness on the order of a few nanometers and a thickness of approximately 0.15 mm. The slide was positioned on a rotational platform. The cone-jet mode of electrospray was established using the experimental setup described above. Then, the droplet was placed below the jet by displacing the slide over the metallic plate with an actuator coupled to a stepper motor (C). After a transient regime of around 100 ms, the cone-jet mode stabilized, and the cells transported by the jet were deposited on the droplet (Fig.\ \ref{Exp_Sketch2}). Cell-by-cell deposition was visualized using the image acquisition system described above. The results presented in Sec. \ref{sec31b} were obtained for $H=1$ mm, $Q=0.05$ ml/h, and $V=3.6$ kV. These values practically correspond to the same electric field $V/H$ as those of the experiments in Sec.\ \ref{sec31}. The distance $H$ was increased to accommodate the droplet. 

\subsection{Cell proliferation and viability assays}

We conducted a cell proliferation assay to evaluate cellular recovery after contact with PEG 35 kDa (10\% w/v). As it was not possible to ensure sterile conditions for long-term cell studies using the electrospray setup, a simulation of the osmotic shock experienced by the cells during the process was performed under sterile conditions in a laminar flow hood. MCF-7 cells were mixed with the PEG 35 kDa at a density of $2\times 10^6$ cells/ml under sterile conditions. Subsequently, 1.66 $\mu$l/min of this suspension was transferred for 30 min into a tube containing high-glucose DMEM GlutaMAX, thereby simulating the contact times with PEG 35 kDa and the culture medium during the electrospray process. After this period, cells were seeded into a 96-well plate at a density of 10,000 cells per well. Cell viability was assessed in triplicate at 24, 48, and 72 h using the Celltiter 96\copyright AQueous One Solution cell proliferation assay (MTS, Promega), following the manufacturer’s instructions. Absorbance values were normalized to cells not exposed to PEG 35 kDa, which were considered as 100\% viability. The results presented in Sec.\ \ref{sec32} are expressed as mean $\pm$ standard deviation ($n=3$). Comparisons between groups were performed using a t-test, considering differences statistically significant for $p<0.05$.

The assay of the cell viability after electrospray was conducted as follows. Cells were mixed with PEG 35 kDa (10\% w/v) at a cell density of $2\times 10^6$ cells/ml. The cell suspension was processed by electrospray and deposited directly into a reservoir containing 2 ml of high-glucose DMEM GlutaMAX. As a control, 1.66 $\mu$l/min of the cell suspension in PEG 35 kDa was transferred for 30 min into a tube containing 2 ml of high-glucose DMEM GlutaMAX. After deposition, cells collected in the reservoir were centrifuged and transferred to a well of an $\mu$-Slide 8 Well chambered coverslip (Ibidi GmbH, Martinsried, Germany). Cells corresponding to the control condition were seeded in an additional well. Cells were stained with the nuclear dye Hoechst 33342 (Thermo Fisher Scientific; 5 $\mu$g/ml) and the viability dye propidium iodide (PI; Thermo Fisher Scientific; 10 $\mu$g/ml), followed by incubation for 30 min at 37 $^{\circ}$C and 5\% CO$_2$. Samples were subsequently analyzed by confocal microscopy (Olympus Fluoview FV1000; 10× objective). The images were analyzed with Fiji\copyright. Comparisons between groups were performed using a t-test, considering $p<0.05$ as statistically significant.

\section{Results}

\subsection{Cell ejection in the steady cone-jet mode}
\label{sec31}

First, we analyzed cell ejection in the steady cone-jet mode of electrospray. This mode was established after a short transient phase of approximately 100 ms. Electric charges accumulated at the Taylor cone surface so that the electric field practically vanished in the cone. The Maxwell stress at the interface accelerated the liquid in the critical cone-jet transition region. A jet of diameter $d_j\simeq 3$ $\mu$m and velocity $v_j\simeq 4$ m/s tapered from the cone tip. These values differ from those predicted by the scaling laws $d_j\sim \left[\rho \varepsilon_o Q^3/(\sigma K)\right]^{1/6}\simeq 0.9$ $\mu$m and $v_j\sim \left[\sigma K/(\rho \varepsilon_o)\right]^{1/3}\simeq 34$ m/s \citep{G04a,GLHRM18} for a Newtonian liquid. It is worth noting that the extensional rate can reach values larger than the inverse of the liquid extensional relaxation time $\lambda_s$ in the cone-jet transition region, which significantly increases the extensional viscosity \citep{M24}. This may explain the observed deviations from predictions based on the Newtonian scaling laws.

\begin{figure}
\begin{center}
\resizebox{0.3\textwidth}{!}{\includegraphics{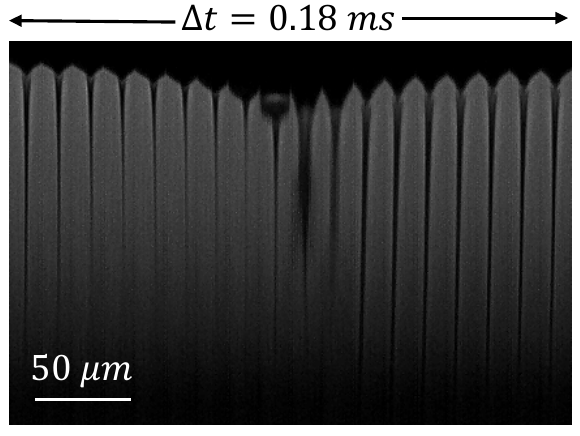}}
\end{center}
\vspace{0cm}
\caption{Sequence of images of the cone tip showing the steady cone-jet mode re-stabilization after cell ejection.}
\label{exp2b}
\end{figure}

The cells moved throughout the cone, dragged by the liquid, with a velocity much smaller than that of the jet. They eventually reached the cone-jet transition region (Fig.\ \ref{exp2}) and escaped from the liquid cone toward the chamber at a speed lower than $v_j$, as shown below. The liquid meniscus stretched and destabilized during the cell ejection event. Then, it re-stabilized within approximately 0.18 ms (Fig.\ \ref{exp2b}). This time is commensurate with the inertio-capillary time $t_{ic}=(\rho R_o^3/\sigma)^{1/2}\simeq 0.25$ ms, where we have chosen the outer radius $R_o=D_o/2$ of the feeding capillary as the liquid meniscus characteristic length. This time is much shorter than the average time lapse between the ejection of two consecutive cells, as shown below. In this sense, one can state that the ejection is ``instantaneous".

\begin{figure}
\begin{center}
\resizebox{0.5\textwidth}{!}{\includegraphics{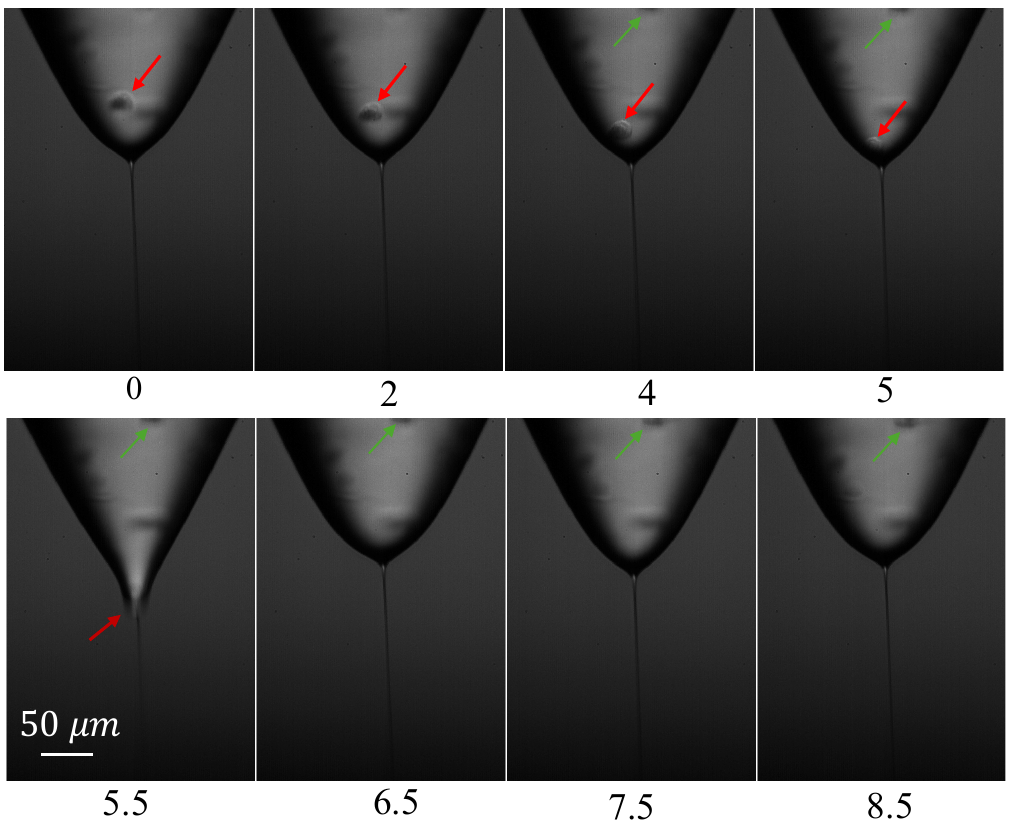}}
\end{center}
\vspace{0cm}
\caption{Electrospray liquid cone. The labels indicate the time measured in milliseconds. The arrows point to the cells.}
\label{exp2}
\end{figure}

The jet was much thinner than the cells, allowing individual cells to be visualized. Our technique achieved the maximum possible spatial resolution, i.e., the cell size. The jet transporting the cell is reduced to a thin film that sheathes the cell from the electric field (Fig.\ \ref{exp2}). The jet diameter is 30 times narrower than the usual one obtained by extrusion ($\sim 100$ $\mu$m), which is insufficient for single-cell printing \citep{B21b,ZFL023,KAPFGLXS24,GFLH20,DRGPSEO25}.

\begin{figure}
\begin{center}
\resizebox{0.4\textwidth}{!}{\includegraphics{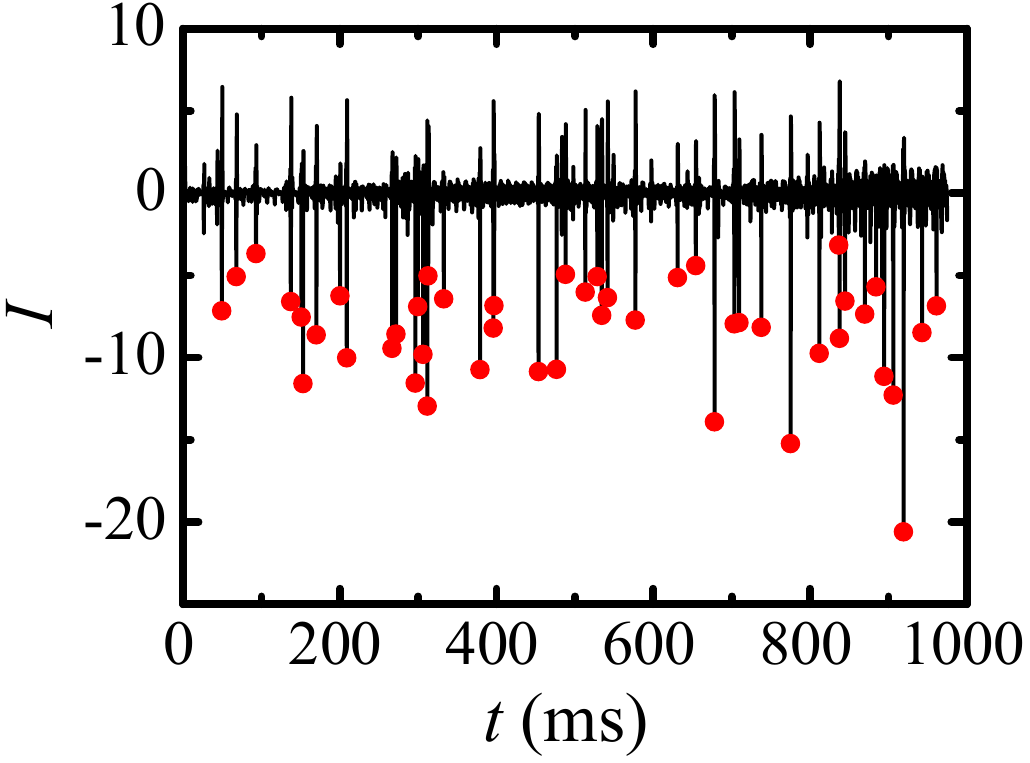}}
\end{center}
\vspace{0cm}
\caption{Normalized total grey intensity $I$ of the images acquired during the ejection of cells. $I=0$ corresponds to the mean grey intensity of all the images. The red circles indicate the instants at which the cells were ejected.}
\label{count}
\end{figure}

Cells can be counted in our experiments using simple image analysis. Figure \ref{count} shows the total grey intensity level $I$ (the sum of the grey intensity level of all the pixels) of an image versus the instant at which it was taken. The peaks indicated by the red circles correspond to a single cell ejection, and were produced by the Taylor cone stretching during the cell ejection (Fig.\ \ref{exp2}). The grey intensity peaks were detected using a custom-made MATLAB script, which identified local minima in the normalized grey-intensity time series. 

The average ejection frequency was $f=49.3$ cells/s, approximately equal to the expected value $f=c\, Q=55.6$ cells/s. This indicates that cells were convected by the liquid toward the cone tip and did not accumulate or become trapped in the injection line. The average time lapse $\Delta t$ between the ejection of two consecutive cells was $\Delta t=1/f=20.3$ ms, and the standard deviation was 15.1 ms, which indicates the degree of periodicity of the cell ejection. As anticipated above, $\Delta t$ is much larger than the time for the cone-jet mode re-stabilization after the cell ejection, implying that the ejection can be regarded ``instantaneous". 

\subsection{Single cell deposition}
\label{sec31b}

We now illustrate the applicability of our technique by depositing cells onto a millimeter-sized droplet of cell culture medium, following the procedure described in Sec.\ \ref{sec21}. For this purpose, we operate very close to the minimum-flow-rate stability limit \citep{GRM13,M24}. In fact, the flow rate $Q=0.05$ ml/h in this experiment is commensurate with the scaling law prediction for the minimum value $Q_{\textin{min}}=\sigma \varepsilon/(\rho K)\simeq 0.024$ ml/h \citep{GRM13,M24}. This entails a fundamental advantage: the time elapsed between the ejection of two consecutive cells, $\Delta t=1/(c\, Q)$, is sufficiently large to place them at distinct locations, even at a relatively high cell concentration $c=2\times 10^6$ cells/ml. 

\begin{figure*}
\begin{center}
\resizebox{0.7\textwidth}{!}{\includegraphics{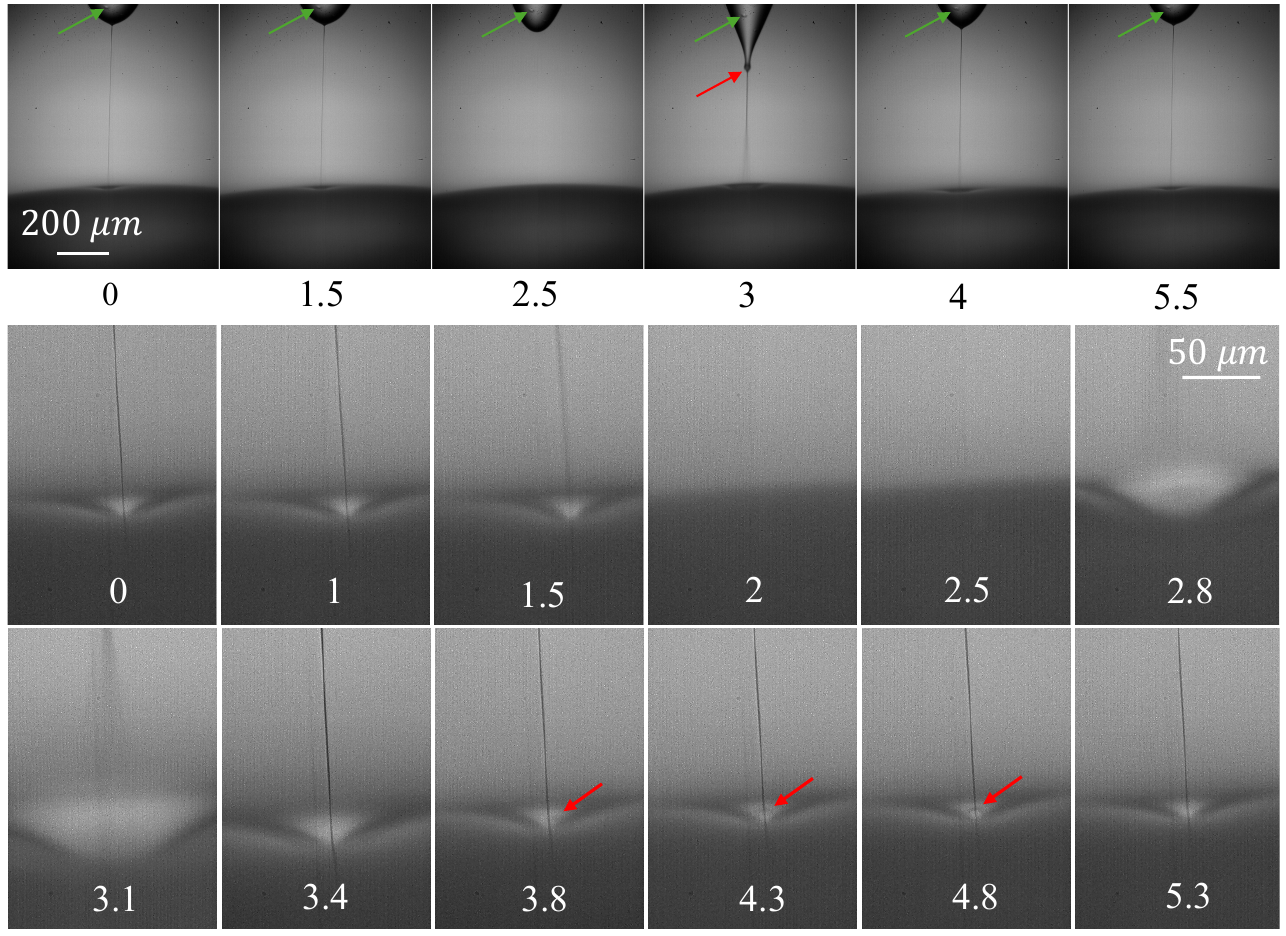}}
\end{center}
\vspace{0cm}
\caption{Sequence of images showing the deposition of a cell onto the droplet of cell culture medium. The labels indicate the time measured in milliseconds.}
\label{dep}
\end{figure*}

The upper images in Fig.\ \ref{dep} show the cell ejection. At $t=0$, the electrospray cone-jet mode is stably running. At $t=2.5$ ms, the jet emission is interrupted by a cell located at the liquid cone tip (the cell cannot be visualized due to optical diffraction). At $t=3$ ms, the cell is expelled from the cone (see the red arrow). The Taylor cone stretches during cell ejection, and the jet undergoes whipping instability. At $t=4$ ms, the cell has reached the droplet of cell culture medium, and the cone-jet mode has been re-established. As shown, the cell covers a distance of 0.5 mm between the liquid cone tip and the droplet in less than 1 ms, implying a velocity exceeding 0.5 m/s (the jet velocity was around 4 m/s). The time required for cone-jet mode re-stabilization is significantly longer than in the ejection experiments described in Sec.\ \ref{sec31}, owing to differences in the electric field configurations. In any case, it is much shorter than the time between the ejection of two consecutive cells. The green arrows point to the cell that will be subsequently ejected. 

The lower images in Fig.\ \ref{dep} show the cell deposition onto the droplet with a larger spatiotemporal resolution. As mentioned above, the cone-jet mode of electrosprayed is established at $t=0$. The jet impact on the droplet surface creates a ``crater" on it. The jet emission is interrupted by a cell located at the cone tip during the interval from $t=2$ to 2.5 ms. The cell contacts the droplet surface at some instant between 2.5 ms and 3.4 ms. The liquid jet produces larger craters at $t=2.8$ ms and 3.1 ms due to its whipping. At $t=3.4$ ms, the cell cannot be visualized because it is submerged in the droplet. It rises to the surface at $t=3.8$ ms due to the capillary forces, as indicated by the red arrow. Finally, it submerges again at $t=5.3$ ms.  

The average time lapse $\Delta t$ between the ejection of two consecutive cells was $\Delta t\simeq 40$ ms in the experiment described above. This time is larger than that required for a robotized system to displace the substrate once the cell ejection is detected, which allows for placing two consecutive cells at distinct user-defined locations. For instance, the point-to-point positioning takes around 20 ms for a distance of 100 $\mu$m with a Miniature Ultrasonic Motor (MUSM) \citep{WDLZCL25}.

\subsection{Cell proliferation and viability assays}
\label{sec32}

\begin{figure*}
    \centering   \includegraphics[width=0.65\textwidth]{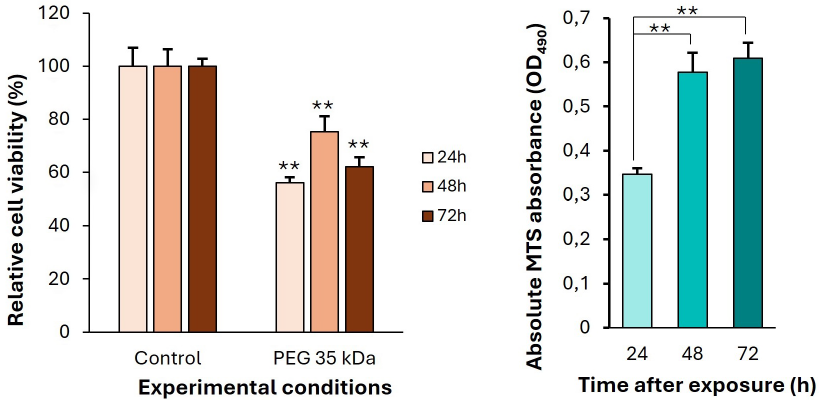}
    \caption{Viability of MCF-7 cells after exposure to PEG 35 kDa (10\% w/v), designed to reproduce the contact times of the electrospray process, as determined by the MTS assay at 24, 48, and 72 h. (A) Relative cell viability normalized to the corresponding control at each time point. (B) Absolute absorbance values of the MTS assay (OD$_{490}$) for cells exposed to PEG 35 kDa. Data are expressed as mean $\pm$ standard deviation ($n=3$). $p<0.01$ compared to the corresponding control (A) or between the indicated time points (B).}
    \label{figc1}
\end{figure*}

The viability of MCF-7 cells after contact with PEG 35 kDa was evaluated using the MTS assay at 24, 48, and 72 h (Figure \ref{figc1}). Our results are expressed relative to the corresponding control at each time point. At 24 h, cell viability was approximately 55\% relative to the control. At 48 h, partial recovery of viability was observed, reaching approximately 75\%, suggesting that a significant fraction of cells recovered after the electrospray process. At 72 h, cell viability decreased slightly relative to the 48 h control, stabilizing at approximately 62\%. The relative reduction in viability observed at 72 h may be attributable to the exponential growth of control cells, which served as the normalization reference, rather than to a delayed cytotoxic effect. In this sense, the proliferation rate of PEG-exposed cells appears lower than that of the control, resulting in lower relative viability values.
 
\begin{figure*}
    \centering   \includegraphics[width=0.75\textwidth]{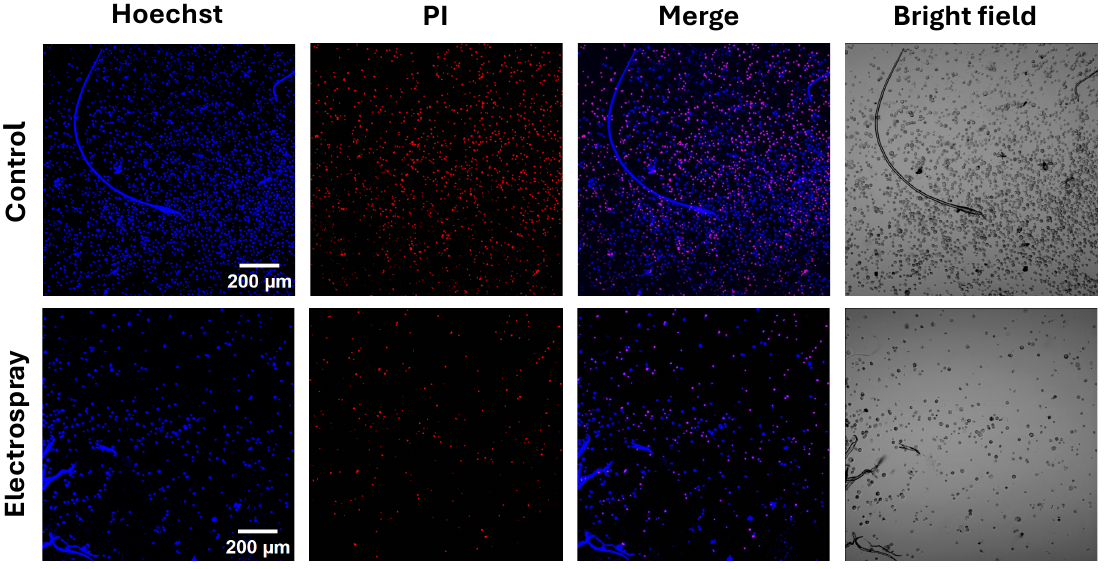}
    \caption{Cell viability analysis by confocal microscopy after the electrospray process. Representative images of control and processed samples showing nuclear staining with Hoechst 33342 (blue), non-viable cells stained with propidium iodide (PI, red), merged images, and bright-field images. Quantitative analysis was performed using three independent fields per condition, and results are expressed as the percentage of viable cells. Data are presented as mean ± standard deviation ($n=3$).}
    \label{figc2}
\end{figure*}

In addition to relative viability normalized to the control, absolute absorbance values from the MTS assay were plotted to evaluate the temporal evolution of the metabolic activity of cells exposed to PEG 35 kDa (Fig.\ \ref{figc2}). These data suggest that viability did not decrease between 48 and 72 h, with comparable absorbance values at both time points, whereas metabolic activity increased significantly between 24 and 48 h. Overall, these results indicate that, following the initial impact of exposure to the PEG solution, cells maintain metabolic activity and reach a stable state from 48 h onward, with no further decline in metabolic activity. This suggests that the surviving cells remain viable in the medium term. 

These results are consistent with previous studies reporting a limited effect of high-molecular-weight PEG on cell viability. In the study by \citet{PNRUFFVVB22}, PEGs of increasing molecular weight (from 200 to 20,000 Da) were dissolved in Phosphate-buffered saline, and MTT assays showed that, after 30 min of exposure, higher-molecular-weight polymers were more effective at preserving cell viability in Caco-2 cells. Similarly, \citet{SZGZYDLXYY24} evaluated the cytotoxicity of PEGs with molecular weights of 600, 2,000, 4,000, and 10,000 Da dissolved in DMEM using an MTT assay, reporting no significant cytotoxicity. In addition, \citet{HBVZ15} confirmed these findings, reporting 88\% viability after incubating Caco-2 cells for 24 h with PEG 35 kDa at 4\% (w/v), applied to the cells under standard culture conditions and assessed by an MTT assay.

In the present study, PEG 35 kDa was dissolved in ultrapure type I water (theoretical osmolality negligible \citep{ZZS25}) to reduce the electrical conductivity. For this reason, the cell suspension was transiently exposed to a hypoosmotic environment with an osmolality far below isotonic conditions ($\sim 300$ mOsmol/kg) \citep{PNRUFFVVB22}. However, the transient nature of the exposure and the subsequent transfer to complete culture medium allowed for recovery of cellular metabolic activity.

To complement the cell proliferation assays and obtain direct information on cellular survival after the electrospray process, a confocal microscopy study was performed (Fig.\ \ref{figc2}). This technique enabled evaluation of the status of deposited cells by distinguishing viable from non-viable cells via fluorescent staining, which allows estimating the percentage of surviving cells after processing. Specifically, cell viability was determined by quantifying Hoechst-stained nuclei and propidium iodide (PI) fluorescence, which marks loss of membrane integrity. In this representative experiment, three independent fields per condition (control and processed samples) were analyzed, with images considered as technical replicates.

As shown in Fig.\ \ref{figc2}, samples subjected to the electrospray process exhibited a lower percentage of viable cells compared to the control. Quantitative analysis revealed an average viable cell percentage of $40.10\pm 6.21$\% in the control and 31.26$\pm 9.84$\% in the processed samples. Although this difference was not statistically significant (t-test, $p>0.05$), a consistent trend toward reduced cell survival after processing was observed across all analyzed fields.

Despite this moderate reduction in cell viability, the results indicate that a substantial fraction of cells retains membrane integrity after contact with the PEG solution and the electrospray process, suggesting that the induced damage is not predominantly irreversible. This observation is particularly relevant, as the preservation of a viable cell population is essential for subsequent proliferation and culture recovery. These findings are consistent with the metabolic assays described above, in which an initial impact associated with exposure to PEG 35 kDa (10\% w/v), characterized by a markedly hypoosmotic environment, was followed by partial recovery of cellular activity after restoration of complete culture medium. Unlike MTS assays, which reflect global metabolic activity, confocal microscopy enables identification of irreversible cell damage at the individual-cell level, thereby providing complementary insight into immediate cell survival after processing.

The viability observed in this study is consistent with previous reports demonstrating that cells can survive and proliferate following electrohydrodynamic deposition. \citet{JT06} reported the injection of a highly concentrated cell suspension of 1321N1 cells (human astrocytoma cell line) using a coaxial nozzle, generating a stable jet that produced droplets with diameters below 50 $\mu$m. After seven days of culture, cells proliferated and reached confluence, demonstrating long-term viability. Similarly, \citet{OIMJ07} successfully deposited RASMC (rabbit aorta smooth muscle cells) and PVSMC (porcine vascular smooth muscle cells) using a coaxial nozzle, encapsulating the cell suspension within an external low-conductivity PDMS phase and maintaining cell survival for at least seven days of culture.

Nevertheless, although both studies achieved long-term cell survival, neither successfully deposited viable cells using a single needle operating under stable jet conditions. In this context, the results of the present work, which suggest appreciable cell survival even after deposition using a single needle, highlight the potential of the proposed system and underscore the importance of optimizing process parameters to minimize impact on cell viability.

\section{Conclusions}

In this study, we present a single-cell deposition technique based on the cone-jet mode of electrospray operated near its minimum stable flow rate. Because the jet diameter is much smaller than that of the cells, individual cells can be clearly observed during deposition, allowing precise control of the number of cells deposited. Operating at such low flow rates allows cells to be positioned at distinct, user-defined locations, even at relatively high cell concentrations. 

Suppose, for instance, that the cells are deposited onto droplets or millimeter wells of cell culture medium, which are placed on a mobile platform. Because of the low value of the flow rate, $Q=0.05$ ml/h, the platform can be moved during the lapse $\Delta t\simeq 40$ ms between the ejection of two consecutive cells, even for a moderately large cell concentration $c=2\times 10^6$ cells/ml. Cells can be easily detected with an optical trigger or by measuring the electric current between the two electrodes (the cell disrupts the continuous transport of electric charge, which can be detected with a picoammeter). Thus, the ejection can be readily synchronized with the robotic platform, enabling each cell to be deposited into a different droplet or well. This can be useful for several techniques, such as single-cell cloning (which produces a pure clone from a single parental cell \cite{YS18}), proteomics \cite{GMHSACT23}, transcriptomics \cite{YMZT25}, or genomics \cite{HXYL23}. We demonstrate the performance of this method by depositing cells one by one onto a millimeter-scale droplet of a standard cell culture medium. 

As mentioned above, our technique enables the manipulation of cells individually, even at a moderately high cell concentration $c=2\times 10^6$ cells/ml, thereby considerably reducing the liquid volume per cell. Droplet microfluidics-based strategies typically operate in the range of $20-60$ nl per droplet \cite{NHSGSY21}, and can reach volumes per cell as low as 10 pl for high-sensitivity single-cell assays. However, this is achieved under Poisson-type statistical encapsulation conditions and using oil/water biphasic systems \cite{LMCTGWP23,ZLFFYCHL25}. Similarly, advanced microfluidic devices enable single-cell confinement in volumes on the order of picoliters or even femtoliters within closed microstructures, thereby maximizing local analyte concentration at the expense of increased architectural complexity and limited integration with open substrates \cite{KMLL18}. In contrast, the proposed system releases, on average, 0.5 nl of liquid between consecutive cells. Although this volume is higher than that of the above-mentioned techniques, operating in continuous-jet mode and in an open architecture allows the elimination of biphasic encapsulation, reduces dependence on confined microchannels, and creates favorable conditions for direct integration with functional surfaces or biofabrication platforms. 

When the cell recipient medium is a tissue, our technique also enables single-cell-level spatial resolution, offering advantages over other high-spatial resolution techniques such as contact \citep{QWHLCZ21}, laser-based \citep{AGENES21,BGKMMLL24}, inkjet, and photopolymerization bioprinting \citep{GCL24}.

We evaluated long-term cellular recovery after contact with PEG 35 kDa (10\% w/v) using a laminar flow hood under sterile conditions. After the initial exposure to the PEG solution, cells continue to exhibit metabolic activity and reach a stable state within 48 hours, with no further reduction thereafter. This indicates that the surviving cells remain viable over the medium term. Our cell viability analysis shows that a significant proportion of cells preserve membrane integrity following contact with the PEG solution and throughout electrospray, suggesting that the induced damage is largely reversible. Complementing this study, we also observed a moderate reduction in cell viability attributable to electrospray.

Our study shows that steady cone-jet electrospray can represent an advantageous compromise between small-volume use, individualized cell deposition, temporal control, and integration flexibility, making our technique a robust alternative to conventional microfluidic platforms in applications where direct deposition and scalability are determining factors. Future studies will enable direct evaluation of the effect of the electrospray process on cell proliferation and viability over the medium- and long-term, under sterile conditions, by integrating the electrospray system into a controlled laminar-flow environment. These approaches will enable more precise characterization of cell proliferation kinetics after processing and validation of results obtained from exposure-simulation assays, thereby reinforcing the applicability of electrospray techniques in cell manipulation and biofabrication systems.

\vspace{1cm}

The data that support the findings of this study are available from the corresponding author upon reasonable request. 
\subsection*{Conflict of interest}

The authors declare that they have no conflict of interest.

\section*{Acknowledgement}
This work was financially supported by the Spanish Ministry of Science, Innovation and Universities (grant no. PID2022-140951OB/AEI/10.13039/501100011033/FEDER, UE).  DF-M acknowledges grant PREP2022-000205 funded by MICIU/AEI /10.13039/501100011033 and ESF+. LM-C and JR-R acknowledge funding from project BIOIMP\_ACE\_MAS\_6\_E, co-funded by the European Union through the Interreg VI-A Spain-Portugal Programme (POCTEP) 2021-2027. 

%

\end{document}